\documentclass[11pt,twoside]{article}
\usepackage{FUSE2004}
\usepackage{natbib}

\usepackage{epsf}
\usepackage{psfig}
\usepackage{lscape}

\begin{document}
\title{FUSE Observations of $\eta$ Carinae}
\author{R. C. Iping, G. Sonneborn, T. R. Gull}
\affil{NASA's GSFC, Code 681, Greenbelt, MD 20771}

\begin{abstract}
$\eta$ Carinae was observed by FUSE through the LWRS (30\arcsec x30\arcsec) and HIRS (1.25\arcsec x20\arcsec) apertures in March and April 2004. There are significant differences between the two spectra. About half of the LWRS flux appears to be due to two B-type stars near the edge of the LWRS aperture, 14\arcsec from $\eta$ Car.  
The HIRS spectrum (LiF1 channel) therefore reveals the intrinsic FUV spectrum of $\eta$ Car without this stellar contamination.  The HIRS spectrum contains strong interstellar H$_2$ having high rotational excitation (up to $J=8$).  Most of the atomic species with prominent ISM features (C II, Fe II, Ar I, P II, etc) also have strong blue-shifted absorption to $v\sim -580$ km/s that is associated with expanding debris from the 1840 eruption.  

\end{abstract}

\section{$\eta$ Carinae Background}

The luminous blue variable (LBV) $\eta$ Carinae (HD 93308), a highly evolved, massive and luminous star, shows strong evidence for a binary companion with a 5.5 year periodicity in X-rays, optical and UV (Daminelli 1996, van Genderen et al. 2003, Whitelock et al. 2004, Corcoran et al. 2001).
Most of its luminosity is emitted in the infra-red, but there is flux and variability at all wavelengths from hard X-rays to radio.
The observed double-lobed structure consists of ejecta from the massive outburst in 1840. The orbital parameters of $\eta$ Car are not well constrained, although extensive modelling is on its way (Corcoran et al. 2001; Pittard  \& Corcoran 2002; Corcoran et al. 2005, in preparation).

\begin{figure}
\plotone{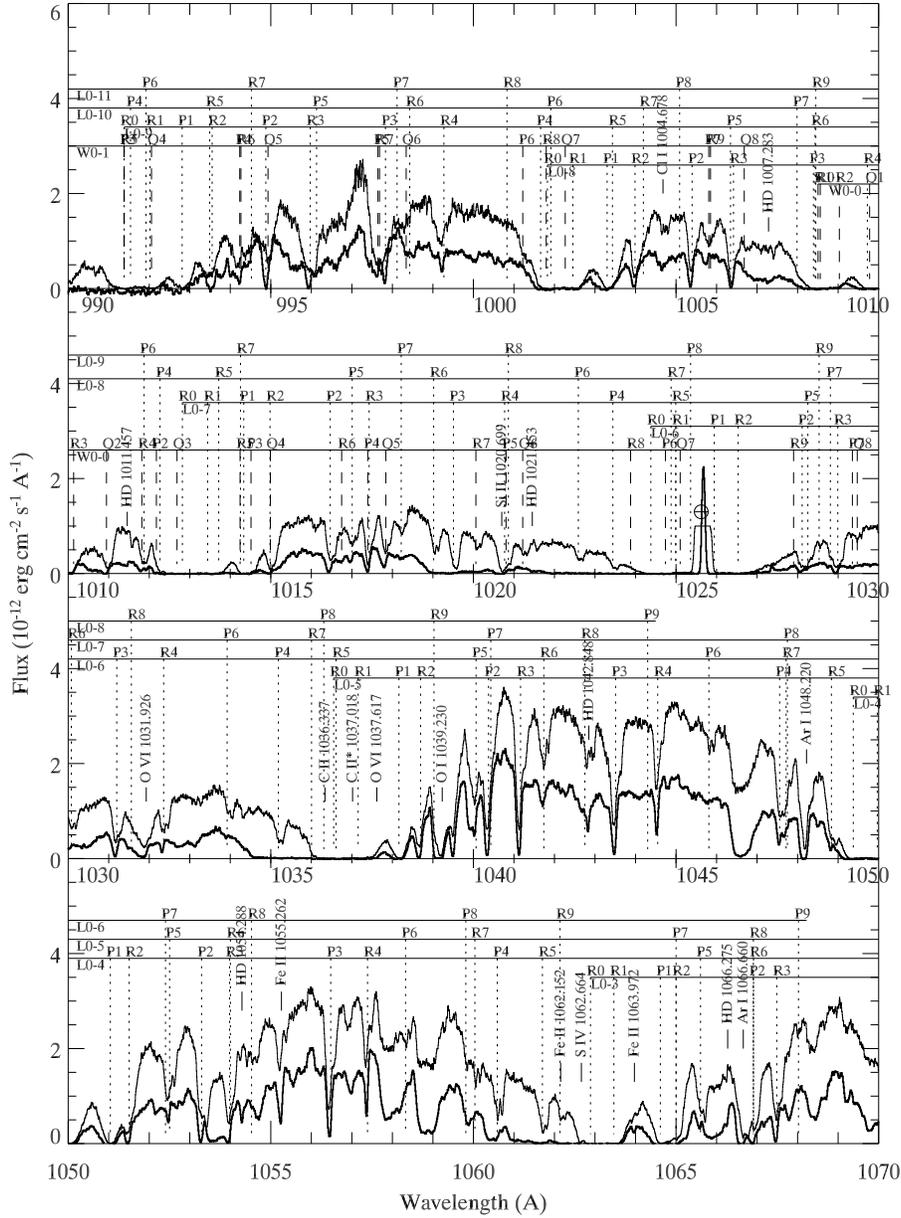}
\caption{The LiF1a spectrum of $\eta$ Car from 990 \AA\ to 1070 \AA.  The thick line of the HIRS aperture spectrum; the thin line is LWRS.  The HIRS flux was scaled up by 1.67X to correct for the point source throughput (60\%). The H$_2$ Lyman and Werner ground state transitions are shown, respectively, by short- and long-dashed lines and are identified by the rotational branch and quantum number (e.g. R3) above each line and by the vibrational quantum numbers (e.g. 0-8) on the left-hand side of each panel.  The principal atomic transitions and the R0 lines of HD are also identified.  All transitions are marked at their laboratory wavelengths.}
\end{figure}

\section{The FUSE Spectrum}
In this paper we report on observations for our FUSE  Cycle 4 GI program D007 obtained in early 2004.  $\eta$ Car was observed through the LWRS aperture (30\arcsec$\times$30\arcsec) on Mar. 29-20 (p.a. $\sim$120\deg) and the HIRS aperture (1\farcs25$\times$20\arcsec) on Apr. 9-11 (p.a. $\sim$ 131\deg).  The exposure times were 41 ksec (LWRS) and 83 ksec (HIRS), yielding a high S/N spectrum, part of which is shown in Figure 1.  The observed LWRS flux at 1040 \AA\ is $\sim3 \times 10^{-12}$
 erg cm$^{-2}$ sec$^{-1}$ \AA$^{-1}$.  The flux level declines
toward the Lyman limit where converging H$_2$  and H I features
completely blanket the spectrum. In the HIRS aperture, after correcting for the 60\% point-source throughput, the 1040\AA\ flux is only $1.5\times 10^{-12}$.  The flux difference appears to be due to two 10$^{th}$-magnitude B-type stars, Tr16-64 and Tr16-65, that are 14\arcsec\ from $\eta$ Car and fall just inside the LWRS aperture.  The inferred FUV flux for these stars ($\sim 1.5\times 10^{-12}$) is consistent with their brightness in HST/ACS 2200\AA\, 2500\AA\, and 3300\AA\ images obtained in 2003 (Smith et al. 2004).  The 
throughput-corrected HIRS FUV flux from $\eta$ Car at 1150-1180\AA\ is 10X higher than that observed in HST/STIS E140M spectra (0\farcs2$\times$0\farcs2 aperture) at the same wavelengths, demonstrating that there is a considerable circumstellar/scattered component to the observed FUV flux.

Numerous broad features ($\sim1$ \AA) in Fig. 1 and 2 are identified as 
absorption by a
forest of high-velocity narrow lines formed in the expanding
circumstellar debris of $\eta$ Car. These
features span a heliocentric velocity range of -140 to -580 km s$^{-1}$
and are seen prominently in low-ionization ground-state
transitions (e.g. N I, Fe II, P II, Ar I, C I, C III 1175-6, and several
unidentified features (see Fig. 2).  The high-velocity components are well-resolved in HST/STIS echelle spectra (Nielsen et al. 2005, ApJS in press).

\begin{figure}
\plotfiddle{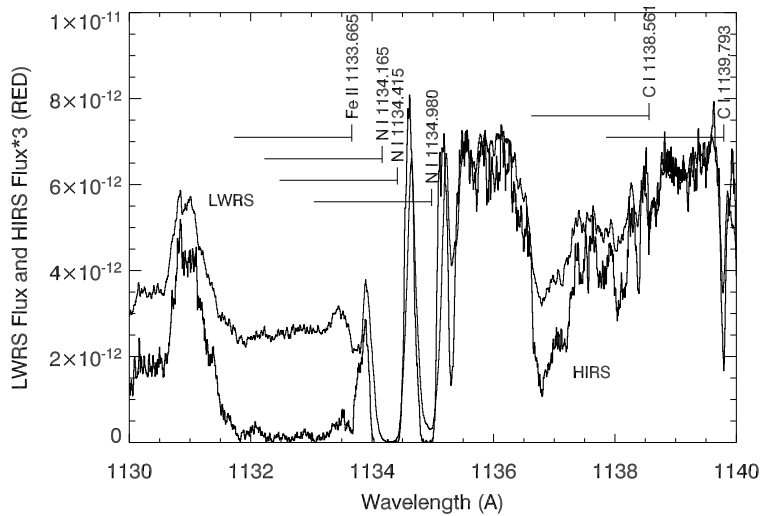}{2.7in}{0}{60}{60}{-190}{-230}
\caption{1130-1140\AA\ portion of  $\eta$ Car LWRS spectrum (thin) and HIRS spectrum (thick). Note
the large difference in Fe II \& N I profiles from 1113-1133\AA. This is due  to two B-stars just
within the LWRS that are excluded by the HIRS aperture. The horizontal lines indicate 0 to $-585$ km s$^{-1}$ and mark the absorption by circumstellar debris from the eruptions in the 1800s. The absorption profiles of C I
1138.56, 1139.79\AA\ also indicate ejecta absorption.}
\end{figure}

\acknowledgements 

These observations were obtained through the FUSE GI program D007.  We are indebted to the positive support of B-G Andersson and the FUSE mission planning team.




\end{document}